\documentclass[preprint,preprintnumbers,showpacs,amsmath,amssymb]{revtex4-1}
\usepackage{graphicx}
\usepackage{epsfig}
\usepackage{amssymb}
\usepackage{graphicx}
\usepackage{dcolumn}
\usepackage{bm}
\newcommand{\ba}{\begin{array}}
\newcommand{\ea}{\end{array}}
\def\br{\begin{eqnarray}}
\def\er{\end{eqnarray}}
\def\be{\begin{equation}}
\def\ee{\end{equation}}

\def\({\left(}
\def\){\right)}

\begin{document}

\title{Energy-dependent dipole form factor in a QCD-inspired model}

\author{C.A.S. Bahia, M. Broilo, and E.G.S. Luna}
\affiliation{Instituto de F\'isica, Universidade Federal do Rio Grande do Sul, Caixa Postal 15051, 91501-970, Porto
Alegre, RS, Brazil }  
 

\begin{abstract}
We consider the effect of an energy-dependent dipole form factor in the high-energy behavior of the forward amplitude. The connection between the semihard parton-level dynamics and
the hadron-hadron scattering is established by an eikonal QCD-based model. Our results for the proton-proton ($pp$) and antiproton-proton ($\bar{p}p$) total cross sections,
$\sigma_{tot}^{pp,\bar{p}p}(s)$, obtained using the CTEQ6L1 parton distribution function, are consistent with the recent data from the TOTEM experiment.
\end{abstract}

\maketitle

\section{Introduction}

The total cross section $\sigma_{tot}(s)$ is a fundamental quantity in collisions of strongly interacting particles. At present one of the main theoretical approaches for the
description of $\sigma_{tot}$ is the QCD-inspired formalism \cite{durand,luna001,luna009,giulia001}. In this approach the energy dependence of the total cross section
is obtained from QCD via an eikonal formulation.
The high-energy dependence of $\sigma_{tot}$ is driven by the rapid increase in gluon density at small-$x$. In this work we explore the
non-perturbative dynamics of QCD in order to describe, in both $pp$ and $\bar{p}p$ channels, the total cross sections $\sigma_{tot}^{pp,\bar{p}p}(s)$ and the ratios of the real to
imaginary part of the forward scattering amplitude, $\rho^{pp,\bar{p}p}(s)$. In our calculations we introduce an energy-dependent dipole form factor which represents the overlap
density for the partons at impact parameter $b$.
The behavior of these forward quantities is derived from the QCD parton model using standard parton-parton elementary processes and an updated set of gluon distribution,
namely the CTEQ6L1 set. In our analysis we have included recent measurements of $pp$ total cross sections at the LHC by the TOTEM Collaboration \cite{TOTEM}. This work is organized
as follows: in the next section we introduce our QCD-based model where the onset of the dominance of semihard partons is managed by the dynamical gluon mass. We investigate a form
factor which the spatial distribution of gluons changes with the energy and introduce integral dispersion relations to connect the real and imaginary parts of eikonals. In
Section III we present our results and in Section IV we draw our conclusions.

\section{The QCD-inspired eikonal model}

In our model the increase of $\sigma_{tot}^{pp,\bar{p}p}(s)$ is associated with elementary semihard processes in the hadrons. In the eikonal representation $\sigma_{tot}(s)$
and $\rho(s)$ are given by
\begin{eqnarray}
\sigma_{tot}(s)   =  4\pi   \int_{_{0}}^{^{\infty}}   \!\!  b\,   db\,
[1-e^{-\chi_{_{R}}(s,b)}\cos \chi_{_{I}}(s,b)]
\label{eq01}
\end{eqnarray}
and
\begin{eqnarray}
\rho(s) = \frac{-\int_{_{0}}^{^{\infty}}   \!\!  b\,  
db\, e^{-\chi_{_{R}}(s,b)}\sin \chi_{_{I}}(s,b)}{\int_{_{0}}^{^{\infty}}   \!\!  b\,  
db\,[1-e^{-\chi_{_{R}}(s,b)}\cos \chi_{_{I}}(s,b)]} ,
\label{eq03}
\end{eqnarray}
respectively, where $s$ is the square of the total center-of-mass energy, $b$ is the impact parameter, and
$\chi(s,b)=\textnormal{Re}\, \chi(s,b) + i\textnormal{Im}\, \chi(s,b) \equiv\chi_{_{R}}(s,b)+i\chi_{_{I}}(s,b)$ is the (complex) eikonal function. The eikonal functions for $pp$ and
$\bar{p}p$ scatterings are the sum of the soft and semihard (SH) parton interactions in the hadron-hadron collision,
\begin{eqnarray}
\chi(s,b) = \chi_{_{soft}}(s,b) + \chi_{_{SH}}(s,b).
\end{eqnarray}
It follows from the QCD parton model that the real part of the eikonal, $\textnormal{Re}\,\chi_{_{SH}}(s,b)$, can be factored as
\begin{eqnarray}
\textnormal{Re}\,\chi_{_{SH}}(s,b) = \frac{1}{2}\, W_{\!\!_{SH}}(b)\,\sigma_{_{QCD}}(s),
\end{eqnarray}
where $W_{\!\!_{SH}}(b)$ is an overlap density for the partons at impact parameter space $b$,
\begin{eqnarray}
W_{\!\!_{SH}}(b) &=& \int d^{2}b'\, \rho_{A}(|{\bf b}-{\bf b}'|)\, \rho_{B}(b'),
\label{ref009}
\end{eqnarray}
and $\sigma_{_{QCD}}(s)$ is the usual QCD cross section 
\begin{eqnarray}
\sigma_{_{QCD}}(s) &=& \sum_{ij} \frac{1}{1+\delta_{ij}} \int_{0}^{1}\!\!dx_{1}
\int_{0}^{1}\!\!dx_{2} \int_{Q^{2}_{min}}^{\infty}\!\!d|\hat{t}|
\frac{d\hat{\sigma}_{ij}}{d|\hat{t}|}(\hat{s},\hat{t}) \nonumber \\
 &\times & f_{i/A}(x_{1},|\hat{t}|)f_{j/B}(x_{2},|\hat{t}|)\, \Theta \! \left( \frac{\hat{s}}{2} - |\hat{t}| \right),
\label{eq08}
\end{eqnarray}
with $|\hat{t}|\equiv Q^{2}$, $\hat{s}=x_{1}x_{2}s$ and $i,j=q,\bar{q},g$.

The eikonal functions for $pp$ and $\bar{p}p$ scatterings are written as $\chi_{pp}^{\bar{p}p}(s,b) = \chi^{+} (s,b) \pm \chi^{-} (s,b)$, with
$\chi^{+}(s,b) = \chi^{+}_{_{soft}}(s,b) + \chi^{+}_{_{SH}}(s,b)$ and $\chi^{-}(s,b) = \chi^{-}_{_{soft}}(s,b) + \chi^{-}_{_{SH}}(s,b)$. Since in the parton model
$\chi^{-}_{_{SH}}(s,b)$ decreases rapidly with increasing $s$, the difference between $pp$ and $\bar{p}p$ cross sections is due only to the different weighting of the quark-antiquark
(valence) annihilation cross sections in the two channels. Thus the crossing-odd eikonal $\chi^{-}(s,b)$ receives no contribution from semihard processes at high energies, and it
is sufficient to take $\chi_{_{SH}}(s,b)=\chi^{+}_{_{SH}}(s,b)$ and, consequently, $\chi^{-}(s,b) = \chi^{-}_{_{soft}}(s,b)$. The connection between the real and imaginary parts of
the crossing-even eikonal $\chi^{+}(s,b)$ is obtained by means of dispersion relation
\begin{eqnarray}
\textnormal{Im}\,\chi^{+}(s,b) = -\frac{2s}{\pi}\, {\cal P}\!\! \int_{0}^{\infty}ds'\,
\frac{\textnormal{Re}\,\chi^{+}(s',b)}{s^{\prime 2}-s^{2}} ,
\label{idr001}
\end{eqnarray}
valid at $s \gg 1$ GeV$^{2}$. The soft term is given by
\begin{eqnarray}
\chi^{+}_{_{soft}}(s,b) = \frac{1}{2}\, W^{+}_{\!\!_{soft}}(b;\mu^{+}_{_{soft}})\, \left[ A' +\frac{B'}{(s/s_{0})^{\gamma}}\, e^{i\pi\gamma/2}
-i C'\left[ \ln\left(\frac{s}{s_{0}}\right) -i\frac{\pi}{2} \right] \right] ,
\label{soft01}
\end{eqnarray}
where $\sqrt{s_{0}}\equiv 5$ GeV and $A'$, $B'$, $C'$, $\gamma$ and $\mu^{+}_{_{soft}}$ are fitting parameters. The odd eikonal $\chi^{-}(s,b)$ is given by
\begin{eqnarray}
\chi^{-}(s,b) &=& \frac{1}{2}\, W^{-}_{\!\!_{soft}}(b;\mu^{-}_{_{soft}})\,D'\, \frac{e^{-i\pi/4}}{\sqrt{s}},
\label{softminus}
\end{eqnarray}
where $\mu^{-}_{_{soft}}\equiv 0.5$ GeV and $D'$ is also a fitting parameter. 

In the expression (\ref{eq08}) the term $d\hat{\sigma}_{ij}/d|\hat{t}|$ is the differential cross section for $ij$ scattering ($i,j=q,\bar{q},g$). We select parton-parton processes
containing at least one gluon in the initial state, i.e., we consider the processes
\begin{eqnarray}
\frac{d\hat{\sigma}}{d\hat{t}}(gg\to gg)=\frac{9\pi\bar{\alpha}^{2}_{s}}{2\hat{s}^{2}}\left(3 -\frac{\hat{t}\hat{u}}{\hat{s}^{2}}-
\frac{\hat{s}\hat{u}}{\hat{t}^{2}}-\frac{\hat{t}\hat{s}}{\hat{u}^{2}} \right) ,
\label{pp001}
\end{eqnarray}
\begin{eqnarray}
\frac{d\hat{\sigma}}{d\hat{t}}(qg\to qg)=\frac{\pi\bar{\alpha}^{2}_{s}}{\hat{s}^{2}}\, (\hat{s}^{2}+\hat{u}^{2}) \left(
\frac{1}{\hat{t}^{2}}-\frac{4}{9\hat{s}\hat{u}} \right) ,
\label{pp002}
\end{eqnarray}
\begin{eqnarray}
\frac{d\hat{\sigma}}{d\hat{t}}(gg\to \bar{q}q)=\frac{3\pi\bar{\alpha}^{2}_{s}}{8\hat{s}^{2}}\, (\hat{t}^{2}+\hat{u}^{2}) \left(
\frac{4}{9\hat{t}\hat{u}}-\frac{1}{\hat{s}^{2}} \right) ,
\label{pp003}
\end{eqnarray}
where 
\begin{eqnarray} 
\bar{\alpha}_{s} (Q^{2})= \frac{4\pi}{\beta_0 \ln\left[
(Q^{2} + 4M_g^2(Q^{2}) )/\Lambda^2 \right]}, 
\label{eq26}
\end{eqnarray}
\begin{eqnarray}
M^2_g(Q^{2}) =m_g^2 \left[\frac{ \ln
\left(\frac{Q^{2}+4{m_g}^2}{\Lambda ^2}\right) } {
\ln\left(\frac{4{m_g}^2}{\Lambda ^2}\right) }\right]^{- 12/11} ;
\label{mdyna} 
\end{eqnarray}
here $\bar{\alpha}_{s} (Q^{2})$ is a non-perturbative QCD effective charge, obtained by Cornwall through the use of the pinch technique in order to derive a gauge invariant
Schwinger-Dyson equation for the gluon propagator \cite{cornwall}.


In the case of semihard gluons we consider the possibility of a ``broadening'' of the spatial distribution. Our assumption suggests an increase of the average gluon radius
when $\sqrt{s}$ increases, and can be properly implemented using the energy-dependent dipole form factor
\begin{eqnarray}
G^{(d)}_{\!\!_{SH}}(s,k_{\perp};\nu_{\!\!_{SH}})=\left( \frac{\nu_{_{SH}}^{2}}{k_{\perp}^{2}+\nu_{_{SH}}^{2}} \right)^{2},
\end{eqnarray}
where $\nu_{_{SH}}= \nu_{1}-\nu_{2}\ln ( \frac{s}{s_{0}} )$, with $\sqrt{s_{0}}\equiv 5$ GeV. Here $\nu_{1}$ and $\nu_{2}$ are constants to be fitted. From equation (\ref{ref009}) we have
\begin{eqnarray}
W_{\!\!_{SH}}(s,b;\nu_{_{SH}}) &=& \frac{1}{2\pi}\int_{0}^{\infty}dk_{\perp}\, k_{\perp}\, J_{0}(k_{\perp}b)\,[G^{(d)}_{\!\!_{SH}}(s,k_{\perp};\nu_{\!\!_{SH}}]^{2} \nonumber \\
 &=& \frac{\nu^{2}_{_{SH}}}{96\pi} (\nu_{_{SH}} b)^{3} K_{3}(\nu_{_{SH}} b).
\end{eqnarray}
Since semihard interactions dominate at high energies, we consider an energy-dependence only in the case of $W_{\!\!_{SH}}(s,b;\nu_{_{SH}})$. In this way the soft overlap densities
$W^{+}_{\!\!_{soft}}(b;\mu^{+}_{_{soft}})$ and $W^{-}_{\!\!_{soft}}(b;\mu^{-}_{_{soft}})$ are static, i.e., $\mu^{+}_{_{soft}}$ and $\mu^{-}_{_{soft}}$ are not energy dependent.

\section{Results}

First, in order to determine the model parameters, we set the value of the gluon mass scale to $m_{g} = 400$ MeV. This choice for the mass scale is not only
consistent to our LO procedure, but is also the one usually obtained in other calculations of strongly interacting
processes \cite{luna001,luna009,luna010,luna2014}. Next, we carry out a global fit to high-energy forward $pp$ and
$\bar{p}p$ scattering data above $\sqrt{s} = 10$ GeV, namely the total cross section $\sigma_{tot}^{pp,\bar{p}p}$ and
the ratio of the real to imaginary part of the forward scattering amplitude $\rho^{pp,\bar{p}p}$. We use data sets
compiled and analyzed by the Particle Data Group \cite{PDG} as well as the recent data at LHC from the TOTEM
Collaboration, with the statistic and systematic errors added in quadrature. We include in the dataset the first estimate for the $\rho$ parameter made by the TOTEM
Collaboration in their $\rho$-independent measurement at $\sqrt{s}=7$ TeV, namely $\rho^{pp}=0.145\pm0.091$ \cite{TOTEM}.


The values of the fitted parameters are given in the legend of Figure 1. The result $\chi^{2}/DOF = 1.062$ for the fit was obtained
for 154 degrees of freedom. The results of the fit to $\sigma_{tot}$ and $\rho$ for both $pp$ and $\bar{p}p$ channels are
displayed in Figure 1, together with the experimental data. The curves depicted in Figure 1 were all calculated using the cutoff $Q_{min}=1.3$ GeV, the value of the CTEQ6 fixed
initial scale $Q_{0}$.

\begin{figure}[!h]
\centering
\hspace*{-0.2cm}\includegraphics[scale=0.60]{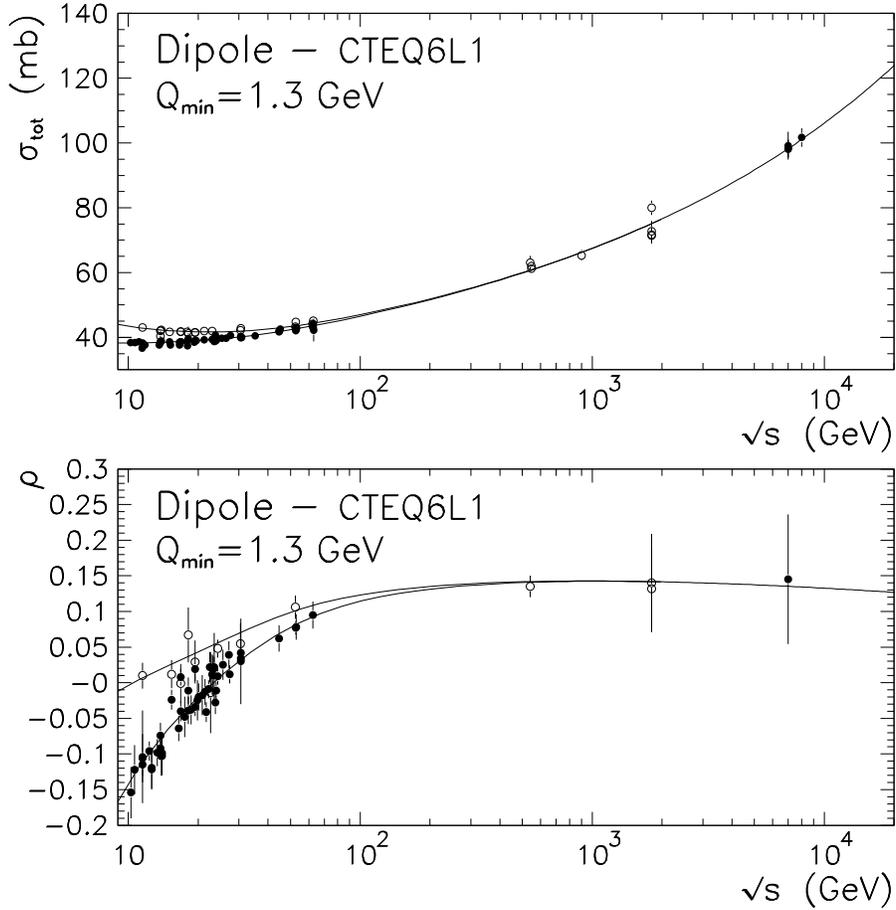}
\label{fig1}
\caption{The total cross section (upper) and the ratio of the real to imaginary part of the forward scattering amplitude (lower), for $pp$ ($\bullet$) and $\bar{p}p$ ($\circ$).
The values of the fitted parameters are: $\nu_{1} = 2.770\pm0.865$ GeV, $\nu_{2} = (7.860\pm5.444)\times 10^{-2}$ GeV, $A' = 108.9\pm8.6$ GeV$^{-1}$, $B' = 30.19\pm15.78$ GeV$^{-1}$,
$C' = 1.260\pm0.437$ GeV$^{-1}$, $\gamma = 0.719\pm 0.200$, $\mu^{+}_{soft} = 0.457\pm0.209$ GeV and $D' = 21.73\pm2.26$ GeV$^{-1}$.}
\end{figure}


\section{Conclusions}

In this work we have investigated infrared effects on semihard parton-parton interactions. We have studied $pp$ and $\bar{p}p$ scattering in the LHC energy region with the
assumption that the increase of their total cross sections arises exclusively from semihard interactions.
In the calculation of $\sigma_{tot}^{pp,\bar{p}p}$ and $\rho^{pp,\bar{p}p}$ we have considered the phenomenological implication of an energy-dependent form factor for semihard
partons. We introduce integral dispersion relations to connect the real and imaginary parts of eikonals with energy-dependent form factors.
In our analysis we have included the recent data at LHC from the TOTEM Collaboration.
Our results show that very good descriptions of $\sigma_{tot}^{pp,\bar{p}p}$ and
$\rho^{pp,\bar{p}p}$ data are obtained by constraining the value of the cutoff at $Q_{min} = 1.3$ GeV.
The $\chi^{2}/DOF$ for the global fit was 1.062 for 154 degrees of freedom. This good
statistical result shows that our eikonal model, where non-perturbative effects are naturally included via a QCD
effective charge, is well suited for detailed predictions of the forward quantities to be measured at higher energies.
In the semihard sector we have considered a form factor in which the average gluon radius increases
with $\sqrt{s}$. With this assumption we have obtained another form in which the eikonal can be factored into the QCD
parton model, namely
$\textnormal{Re}\,\chi_{_{SH}}(s,b) = \frac{1}{2}\, W_{\!\!_{SH}}(s,b)\,\textnormal{Re}\,\sigma_{_{QCD}}(s)$. The imaginary
part of this {\it semi-factorizable} eikonal was obtained by means of appropriate integral dispersion relations
which take into account eikonals with energy-dependent form factors.

At the moment we are analyzing the effects of different updated sets of parton distributions on the
forward quantities, namely CTEQ6L and MSTW, investigating the uncertainty on these forward observables coming from the uncertainties associated with the dynamical
mass scale and the parton distribution functions, and exploring a new ansatz for the energy-dependent form factor.

\smallskip


\begin {thebibliography}{99}

\bibitem{durand} L. Durand and H. Pi, Phys. Rev. Lett. {\bf 58}, 303 (1987);
Phys. Rev. D {\bf 38}, 78 (1988);
{\bf 40}, 1436 (1989).

\bibitem{luna001} E.G.S. Luna, A.F. Martini, M.J. Menon, A. Mihara, and A.A. Natale, Phys. Rev. D {\bf 72}, 034019
(2005);
E.G.S. Luna, Phys. Lett. B {\bf 641}, 171 (2006);
E.G.S. Luna and A.A. Natale, Phys. Rev. D {\bf 73}, 074019 (2006);
E.G.S. Luna, Braz. J. Phys. {\bf 37}, 84 (2007);
D.A. Fagundes, E.G.S. Luna, M.J. Menon, and A.A. Natale, Nucl. Phys. A {\bf 886}, 48 (2012);
E.G.S. Luna and P.C. Beggio, Nucl. Phys. A {\bf 929}, 230 (2014).

\bibitem{luna009} E.G.S. Luna, A.L. dos Santos, and A.A. Natale, Phys. Lett. B {\bf 698}, 52 (2011).

\bibitem{giulia001} A. Corsetti, A. Grau, G. Pancheri, and Y.N. Srivastava, Phys. Lett. B {\bf 382}, 282 (1996);
A. Grau, G. Pancheri, and Y.N. Srivastava, Phys. Rev. D {\bf 60}, 114020 (1999);
R.M. Godbole, A. Grau, G. Pancheri, and Y.N. Srivastava, Phys. Rev. D {\bf 72}, 076001 (2005);
A. Achilli {\it et al.}, Phys. Lett. B {\bf 659}, 137(2008);
A. Grau, R.M. Godbole, G. Pancheri, and Y.N. Srivastava, Phys. Lett. B {\bf 682}, 55 (2009);
G. Pancheri, D.A. Fagundes, A. Grau, S. Pacetti, and Y.N. Srivastava, arXiv:1301.2925 [hep-ph].

\bibitem{TOTEM} G. Antchev {\it et al.}, EPL {\bf 96}, 21002 (2011);
G. Antchev {\it et al.}, EPL {\bf 101}, 21002 (2013);
G. Antchev {\it et al.}, EPL {\bf 101}, 21004 (2013);
G. Antchev {\it et al.}, Phys. Rev. Lett. {\bf 111}, 012001 (2013).

\bibitem{cornwall} J. M. Cornwall, Phys. Rev. D {\bf 22}, 1452 (1980); D {\bf 26}, 1453 (1982).


\bibitem{luna010} A. Doff, E.G.S. Luna, and A.A. Natale, Phys. Rev. D {\bf88}, 055008 (2013).

\bibitem{luna2014} E.G.S. Luna and A.A. Natale, J. Phys. G {\bf 42}, 015003 (2015).

\bibitem{PDG} K.A. Olive {\it et al.}, Chin. Phys. C {\bf 38}, 090001 (2014).



\end{thebibliography}
\end{document}